\documentclass[twocolumn,aps,pr,superscriptaddress,preprintnumbers,nofootinbib,10pt]{revtex4-2}
\usepackage{amsmath}
\usepackage{amssymb}
\usepackage[dvipdf,dvips]{graphicx}
\usepackage{color}
\usepackage{hyperref}
\usepackage{url}
\usepackage{slashed}
\usepackage{subfigure}
\usepackage[usenames,dvipsnames]{xcolor}
\usepackage{amsmath}
\usepackage{amsfonts}
\usepackage{float} 
\usepackage{amssymb}
\usepackage{epsfig}
\usepackage{graphics}
\usepackage{euscript}
\usepackage{slashed}
\usepackage{epstopdf}
\usepackage[utf8]{inputenc}
\allowdisplaybreaks
\usepackage[normalem]{ulem}
\usepackage{pifont}
\usepackage{dsfont}
\usepackage{MnSymbol}
\usepackage{verbatim}
\usepackage{graphicx}
\usepackage{latexsym}
\usepackage{tikz-feynman}

\begin{document}
	
	\title{Mermin's inequalities in Quantum Field Theory}

	\author{Philipe De Fabritiis} \email{pdf321@cbpf.br} \affiliation{CBPF $-$ Centro Brasileiro de Pesquisas Físicas, Rua Dr. Xavier Sigaud 150, 22290-180, Rio de Janeiro, Brazil}

	\author{Itzhak Roditi} \email{roditi@cbpf.br} \affiliation{CBPF $-$ Centro Brasileiro de Pesquisas Físicas, Rua Dr. Xavier Sigaud 150, 22290-180, Rio de Janeiro, Brazil}
	
	\author{Silvio Paolo Sorella} \email{silvio.sorella@gmail.com} \affiliation{UERJ $–$ Universidade do Estado do Rio de Janeiro,	Instituto de Física $–$ Departamento de Física Teórica $–$ Rua São Francisco Xavier 524, 20550-013, Maracanã, Rio de Janeiro, Brazil}
	
	\begin{abstract}
A relativistic Quantum Field Theory framework is devised for Mermin's inequalities. By employing smeared Dirac spinor fields, we are able to introduce unitary operators which create, out of the Minkowski vacuum $\vert 0 \rangle $, GHZ-type states. In this way, we are able to obtain a relation between the expectation value of Mermin's operators in the vacuum and in the GHZ-type states. We show that Mermin's inequalities turn out to be maximally violated when evaluated on these states. 
	\end{abstract}

	\maketitle	
	
	\section{Introduction}

\vspace{-0.1cm}	
	
The phenomenon of entanglement is one of the most striking issues of nowadays physics, leading to an impressive set of developments,  from both theoretical and technological points of view, ranging from the physics of the black holes to the engineering  of the quantum computation. Among the various available proposed  quantities allowing  to  quantify entanglement, it is fair to say that the violation of the Bell-CHSH inequality~\cite{Bell:1964kc,Clauser:1969ny} plays a prominent role due to its historical origins, to its experimental verifications, see~\cite{Clauser:1969ny,Freedman:1972zza,Clauser:1974tg,Clauser:1978ng, aspect0,aspect1,aspect2,aspect3,z0,z1} and Refs. therein, as well as to its relevance for recent subjects that might have important repercussions on the creation of new technologies, such as teleportation and quantum cryptography. 

Besides the extensive research activity dedicated to this subject in Quantum Mechanics, the Bell-CHSH inequality has been also investigated within the realm of relativistic Quantum Field Theory~\cite{Summers:1987fn,Summ,Summers:1987ze,Summers:1988ux,Summers:1995mv,Verch}, see also ~\cite{Peruzzo:2022pwv,Peruzzo:2022tog} for more recent attempts. It is worth emphasizing that this is a topic of great  interest in the phenomenology of Particle Physics, much related to both current and future experiments at the Large Hadron Collider, LHC,  and other particle accelerators, that will make it possible to scrutinize this subject in energy regimes never considered before, as one might check out in Refs.~\cite{top0, top1, top2,top3,Afiknovo,Lambda,tauphotons,eletronpositron,neutralmeson,neutrino,charmonium1,charmonium2, positronium,HiggsW,Ashby-Pickering:2022umy,Fabbrichesi:2023cev}. 

In particular, by using techniques from Algebraic Quantum Field Theory~\cite{Haag:1992hx}, the authors~\cite{Summers:1987fn,Summ,Summers:1987ze,Summers:1988ux,Summers:1995mv} have been able to show that even free quantum fields lead to the maximum violation of the Bell-CHSH inequality, {\it i.e.} to Tsirelson's bound~\cite{tsi1,tsi2}: $2 \sqrt{2}$. This result  can be seen as a confirmation of the severity of entanglement in relativistic Quantum Field Theory. This statement can be justified by pointing out that the Minkowski vacuum state of a Quantum Field Theory is known to be a very entangled state, a feature which can be made manifest by using Rindler wedges geometry, which enable us to express in an explicit way the Minkowski vacuum in terms of right and left Rindler modes~\cite{Crispino:2007eb,Harlow:2014yka,Donnelly:2015hxa,Blommaert:2018rsf,Blommaert:2018oue}. 

Therefore, due to the highly complex structure of the Minkowski vacuum, it is expected that particularities of multipartite systems are still present.  Thus, we consider that it is worthy to go beyond the   Quantum Field Theory  investigation of the Bell-CHSH inequality in order to include other kinds of inequalities which are largely employed in  Quantum Mechanics. This is the case of Mermin's inequalities~\cite{Mermin}.  These inequalities enable one to perform dichotomic  measurements in multipartite systems. As such, Mermin's inequalities have found a large number of applications, see~\cite{Alsina16a,Alsina16b} and references therein, including their experimental verification~\cite{Lanyon14, Zelaquett}. 

The construction of  Mermin's inequalities for a $n$-partite system is done recursively, starting from the lowest order, which is nothing but Bell-CHSH inequality. In this sense, Mermin's inequalities are a generalization of the Bell-CHSH inequality to $n$-partite systems. Interestingly enough, the maximum violation of  Mermin's inequalities can be achieved by making use of the Greenberger–Horne–Zeilinger (GHZ) states~\cite{z0}. 

It is the aim of the present work to provide a formulation of Mermin's inequalities within the framework of relativistic Quantum Field Theory,  allowing us to shed light on the non-local correlations present in the vacuum.

The paper is organized as follows: In Section~\ref{Mineq} we provide a short account of the recursive construction of  Mermin's polynomials in   Quantum Mechanics, and of the violation of the corresponding inequalities by employing the GHZ states. Section~\ref{DF} provides the  smearing procedure details~\cite{Haag:1992hx} for the Dirac spinor field in $1+3$ Minkowski spacetime. As we shall see, the use of spinor fields turns out to be particularly convenient in view of the rather simple and elegant formulation of the GHZ states via fermionic creation and annihilation smeared operators. In Section~\ref{SQU} we discuss how the GHZ states can be obtained by acting on the vacuum state by means of suitable unitary fermionic squeezing operators. In Section~\ref{M3M4} we present the calculations of  Mermin's inequalities violations in   Quantum Field Theory  corresponding to the polynomials $M_3$ and $M_4$. It turns out that both $M_3$ and $2M_4$ attain their maximum violation given, respectively, by $4$ and $8\sqrt{2}$. Section~\ref{conc} contains our conclusions.

\section{Mermin's inequalities and GHZ states in Quantum Mechanics}\label{Mineq}

We provide here a brief summary of Mermin's inequalities. The so-called Mermin's polynomials can be defined in a recursive manner, see Ref.~\cite{Alsina16a}, according to the general rule:
	\begin{align}\label{key}
		M_n = \frac{1}{2} M_{n-1} \left(A_n + A'_n\right) + \frac{1}{2} M'_{n-1} \left(A_n - A'_n\right).
	\end{align} 
In the above expression, the $M_n'$ operators can be obtained from $M_n$ upon changing $A_n \rightarrow A_n'$ and $A_n' \rightarrow A_n$. Here we shall make use of   $M_1 = 2 A_1$ as the first term in the recursive procedure. Adopting this definition, the maximum value one can obtain in   Quantum Mechanics  is given by the following upper bound
	\begin{align}\label{key}
		\vert M_n \vert \leq 2^{\left(\frac{n+1}{2}\right)}.
	\end{align}
For example, Mermin's polynomial for 3-qubits is given by
\begin{align}\label{M3}
		M_3 = A' B C + A B' C + A B C' - A' B' C'.
\end{align}
Considering the absolute value, local realistic theories  obey the bound $\vert \langle M_3 \rangle_{\text{Cl}} \vert \leq 2 $. For   Quantum Mechanics, on the other hand, the upper bound is given by $\vert \langle M_3 \rangle_{\text{QM}}\vert \leq 4$.
Notice that if we choose parameters such that $C = C' = 1$,  Mermin's polynomial $M_3$ reduces to the Bell-CHSH operator, that is,
\begin{align}\label{CHSH}
	\mathcal{C}_{\text{CHSH}} = A B + A' B + A B'   - A' B'.
\end{align}
In this simpler case,  the local realistic bound gives $\langle \mathcal{C}_{\text{CHSH}} \rangle_{\text{Cl}} \leq 2$, while the quantum one is given by $\langle \mathcal{C}_{\text{CHSH}} \rangle_{\text{QM}} \leq 2 \sqrt{2}$, the so-called Tsirelson's bound. 

Proceeding with the recursive expression,  Mermin's polynomial for 4-qubits can be written as
	\begin{align}\label{M4}
		2 M_4 &= -A B C D \nonumber \\
		&+ \left(A' B C D + A B' C D + A B C' D + A B C D'\right) \nonumber \\
		&+ \left( A' B' C D + A' B C' D + A' B C D'   \right. \nonumber \\
		&\left. \,\,\,   + A B' C' D + A B' C D' + A B C' D'      \right) \nonumber \\
		&-\left(A' B' C' D + A' B' C D' + A' B C' D' + A B' C' D'\right) \nonumber \\
		&-A' B' C' D'.
	\end{align}
In absolute values, the classical bound is  $\langle 2 M_4 \rangle_{\text{Cl}} \leq 4$, and the quantum bound is given by $\langle 2 M_4 \rangle_{\text{QM}} \leq 8 \sqrt{2}$. 

To see how the maximum violation of  Mermin's operators $M_3$ and $M_4$ is achieved, we shall consider the Greenberger–Horne–Zeilinger (GHZ) state~\cite{z0}, a well-known entangled state, usually associated with 3 spin 1/2 particles, defined as:
\begin{align}\label{key}
	\vert \text{GHZ} \rangle_3 = \frac{1}{\sqrt{2}} \left(\vert 0, 0, 0 \rangle + \vert 1, 1, 1 \rangle \right),
\end{align}
where $ \vert \cdot, \cdot, \cdot \rangle $  stands for $ \vert \cdot \rangle \otimes \vert \cdot \rangle \otimes \vert \cdot \rangle $. 

Let us introduce operators $A, A', B, B', C, C'$ such that for each of these operators, we have $X^2 = 1$, $X^\dagger = X$, and such that $\left[X,Y\right] = 0$ for any pair of them. These operators act on the basis states as
\begin{align}\label{BellDefABC}
	A \vert 0, x, y \rangle &= e^{i \alpha} \vert 1, x, y \rangle, \quad A \vert 1, x, y \rangle = e^{-i \alpha} \vert 0, x, y \rangle, \nonumber \\
	B \vert x, 0, y \rangle &= e^{i \beta} \vert x, 1, y \rangle, \quad B \vert x, 1, y \rangle = e^{-i \beta} \vert x, 0, y \rangle, \nonumber \\
	C \vert x, y, 0 \rangle &= e^{i \gamma} \vert x, y, 1 \rangle, \quad C \vert x, y, 1 \rangle = e^{-i \gamma} \vert x, y, 0 \rangle,
\end{align}  	
for any value of $x,y$, with similar expressions for the primed operators. The quantities $(\alpha, \beta,\gamma)$ are arbitrary parameters, which can be chosen at the best convenience. Thus, one can consider  Mermin's polynomial $M_3$ defined in Eq.~\eqref{M3}, adopting the operators described above and compute its expectation value on the GHZ state. One finds:
\begin{align}\label{key}
	\langle \text{GHZ} \vert M_3 \vert \text{GHZ} \rangle &= \cos(\alpha' + \beta + \gamma) + \cos(\alpha + \beta' + \gamma) \nonumber \\
	&+ \cos(\alpha + \beta + \gamma') - \cos(\alpha' + \beta' + \gamma').
\end{align}
In order to maximize this result, one can choose the following values for the parameters: $\alpha = 0, \alpha' = \frac{\pi}{2}, \beta = -\frac{\pi}{4}, \beta' = \frac{\pi}{4}, \gamma = - \frac{\pi}{4}, \gamma' = \frac{\pi}{4}$. In this way, the GHZ state provides the maximum violation of  Mermin's third order inequality, given by
\begin{align}\label{key}
	\langle \text{GHZ} \vert M_3 \vert \text{GHZ} \rangle = 4.
\end{align}
In the same vein, one can define the generalized GHZ state as an entangled quantum state of $n>2$ subsystems, given by
\begin{align}\label{key}
	\vert \text{GHZ} \rangle_n = \frac{1}{\sqrt{2}} \left(\vert 0, 0, 0..., 0 \rangle_n + \vert 1, 1, ..., 1 \rangle_n \right),
\end{align} 
where $\vert 0, 0, ..., 0\rangle_n$ is understood as the tensor product of $n$ states $\vert 0 \rangle$, and similarly for $\vert 1, 1, ..., 1 \rangle_n$.
For instance, one can consider the case of 4 spin 1/2 particles, thus taking $n=4$, and evaluate the expectation  value of the operator $M_4$ defined in Eq.~\eqref{M4} on the state $\vert \text{GHZ} \rangle_4$, adopting the same definition for the operators that we used above for $n=3$, with the obvious extension for the action of the operators $D,D'$. Therefore, computing $\langle M_4 \rangle  \equiv \langle  \text{GHZ} \vert M_4 \vert \text{GHZ} \rangle$, one finds a sum of cosines with the same sign pattern of the  operator $M_3$, that is:
\begin{align}\label{M4GHZQM}
	\langle 2 M_4 \rangle &= -\cos (\alpha +\beta +\gamma +\delta ) + \cos (\alpha +\beta +\gamma +\delta') \nonumber \\
	&+ \cos (\alpha +\beta +\gamma'+\delta ) + \cos (\alpha +\beta +\gamma'+\delta')\nonumber \\
	&+\cos (\alpha +\beta'+\gamma +\delta ) + \cos (\alpha +\beta'+\gamma +\delta') \nonumber \\
	&+\cos (\alpha +\beta'+\gamma'+\delta ) - \cos (\alpha +\beta'+\gamma'+\delta') \nonumber \\
	&+\cos (\alpha'+\beta +\gamma +\delta ) + \cos (\alpha'+\beta +\gamma +\delta') \nonumber \\
	&+\cos (\alpha'+\beta +\gamma'+\delta ) - \cos (\alpha'+\beta +\gamma'+\delta') \nonumber \\
	&+ \cos (\alpha'+\beta'+\gamma +\delta ) - \cos (\alpha'+\beta'+\gamma +\delta') \nonumber \\
	&-\cos (\alpha'+\beta'+\gamma'+\delta )-\cos (\alpha'+\beta'+\gamma'+\delta').
\end{align}

In order to maximize this expression, one can choose the following parameters: $\alpha = 0, \alpha' = \frac{\pi}{2}, \beta = -\frac{\pi}{4}, \beta' = \frac{\pi}{4}, \gamma = - \frac{\pi}{4}, \gamma' = \frac{\pi}{4}, \delta = - \frac{\pi}{4}, \delta' = \frac{\pi}{4}$. In this way, we find the maximum violation of Mermin's fourth order inequality,  
\begin{align}\label{key}
	 \langle  \text{GHZ} \vert \ 2 M_4 \; \vert \text{GHZ} \rangle = 8 \sqrt{2}.
\end{align}
One could continue with this reasoning and consider, for instance, the expectation value of $M_5$ in the GHZ state obtaining, once again, maximum violation: 
\begin{equation} 
\langle GHZ \vert \; 2M_5 \, \vert GHZ \rangle = 16.
\end{equation} 
The reasoning for higher $M_n$ polynomials would naturally follow a similar pattern and give, once more,  maximum violation for the {\it $n$-th} order inequality.

\section{The Quantized Dirac Field}\label{DF}

As already mentioned in the Introduction, in order to construct a relativistic  Quantum Field Theory  setup for Mermin's inequality, we shall employ Dirac spinor fields in (1+3) Minkowski spacetime. As we shall see, this is due to the fact that spinor fields yield an elegant and very helpful formulation of the GHZ states. Employing the conventions of \cite{Itzykson}, for the 
Dirac action we have:
\begin{align}\label{key}
	S = \int \! d^4x \mathcal{L} = \int \! d^4x \,  \bar{\psi} \left(i \gamma^\mu \partial_\mu - m \right) \psi.
\end{align}
In the above expression,  $\gamma^\mu$ are the Dirac matrices, four $4\times4$ matrices satisfying the Clifford algebra $\lbrace \gamma^\mu, \gamma^\nu \rbrace = 2 g ^{\mu \nu} \mathbf{1}$. The field $\psi$ is a Dirac spinor, a four-component complex field transforming under a Lorentz transformation $\Lambda$ with parameters $\omega_{\mu \nu}$ according to the spinor representation,
\begin{align}\label{key}
 \psi'(x) = e^{-\frac{i}{2} \omega_{\mu \nu} S^{\mu \nu}} \psi(\Lambda^{-1} x),
\end{align} 
where $S^{\mu \nu} = \frac{i}{4} \left[\gamma^\mu, \gamma^\nu\right]$. The Dirac conjugate field $\bar{\psi}$ is defined by $\bar{\psi} = \psi^\dagger \gamma^0$. Finally,  $m$ is a mass parameter. To proceed with the canonical quantization, one expands the Dirac field in terms of the c-number plane wave solutions of the Dirac equation as:
\begin{align}\label{DiracExpansion}
	\psi(x) &= \! \int \!\! \frac{d^3k}{(2\pi)^3} \frac{m}{\omega_k} \sum_{s=1,2} \left[ b_s u^{(s)} e^{-i k x} + d_s^{\dagger} v^{(s)} e^{+i k x}  \right], \nonumber \\
	\bar{\psi}(x) &= \! \int \!\! \frac{d^3k}{(2\pi)^3} \frac{m}{\omega_k} \sum_{s=1,2} \left[ b_s^\dagger \bar{u}^{(s)} e^{+i k x} + d_s \bar{v}^{(s)} e^{-i k x}  \right],
\end{align}
where we defined $\omega_k = \sqrt{\vert \vec{k}\vert^2 + m^2}$, and  $b, b^\dagger, d, d^\dagger$ satisfy the following anticommutation relations
 \begin{align}\label{key}
 	\lbrace b_r(k), b_s^\dagger(q) \rbrace &= (2\pi)^3 \frac{k^0}{m} \delta^3(\vec{k} - \vec{q}) \delta_{r s}, \nonumber \\
 	 \lbrace d_r(k), d_s^\dagger(q) \rbrace &= (2\pi)^3 \frac{k^0}{m} \delta^3(\vec{k} - \vec{q}) \delta_{r s},
 \end{align}
and all other anticommutators vanish. Thus, for the equal time canonical anticommutation relations between the Dirac fields, we have:
\begin{align}\label{key}
	\lbrace \psi_\alpha(t, \vec{x}), \psi_\beta^\dagger(t, \vec{y}) \rbrace = \delta^3(\vec{x} - \vec{y}) \delta_{\alpha \beta}.
\end{align}
From the anticommutation relations, one can also obtain the following useful spin sums:
\begin{align}\label{key}
	\sum_{s = 1,2} u^{(s)}_\alpha(k) \bar{u}^{(s)}_\beta(k) &= \left(\frac{\slashed{k} + m}{2m}\right)_{\alpha \beta}, \nonumber \\
	\sum_{s = 1,2} v^{(s)}_\alpha(k) \bar{v}^{(s)}_\beta(k) &= \left(\frac{\slashed{k} - m}{2m}\right)_{\alpha \beta}.
\end{align}

The anticommutator of two Dirac free fields at arbitrary spacetime separation can be obtained from the relations stated above, giving us $\lbrace \psi(x), \psi(y) \rbrace = \lbrace \bar{\psi}(x), \bar{\psi}(y) \rbrace = 0$, as well as: 
\begin{align}\label{key}
	\lbrace \psi_\alpha(x), \bar{\psi}_\beta(y) \rbrace = \left( i \gamma^\mu \partial_\mu + m \right)_{\alpha \beta} i \Delta_{\text{PJ}}(x-y),
\end{align}
where the real distribution $\Delta_{\text{PJ}}(x-y)$ is the well-known Pauli-Jordan distribution, given by the expression:
\begin{align}\label{key}
	\Delta_{\text{PJ}}(x-y) = \frac{1}{i} \int \! \frac{d^4k}{(2 \pi)^3} \delta^4(k^2-m^2) \epsilon(k^0) e^{-i k (x-y)},
\end{align}
where $\epsilon(x) = \frac{x}{\vert x \vert}$. From the above expression, one can immediately see that the Pauli-Jordan distribution is Lorentz-invariant and odd under the change $(x-y) \rightarrow (y-x)$. Moreover, it is a solution of the Klein-Gordon equation and, importantly, vanishes outside of the light cone, encoding the information that measurements at space-like separated points do not interfere, an important consequence of locality and relativistic causality. 

The Fock vacuum of the Minkowski spacetime, $\vert 0 \rangle$, is defined as the state annihilated by all the annihilation operators, that is
\begin{align}\label{key}
	b_1(k) \vert 0 \rangle = b_2(k) \vert 0 \rangle = d_1(k) \vert 0 \rangle = d_2(k) \vert 0 \rangle = 0
\end{align}
Nevertheless, the quantum fields described above are too singular objects, and must be treated as operator-valued distributions in Minkowski spacetime~\cite{Haag:1992hx}. To give a well-defined meaning to these objects, we  define them as tempered distributions that can be smeared out by test functions belonging to the space $\mathcal{C}_0^\infty\left(\mathbb{R}^4\right)$\footnote{The space $\mathcal{C}_0^\infty\left(\mathbb{R}^4\right)$ is  the space of smooth infinitely differentiable functions with compact support.}. Accordingly, we  introduce the smeared Dirac fields, defining them as~\cite{Haag:1992hx}
\begin{align}\label{SmearingDirac}
	\psi(f) &= \int \! d^4x \, \bar{f}(x) \psi(x), \nonumber \\
	\psi^\dagger(f) &= \int \! d^4x  \,\bar{\psi}(x) f(x),
\end{align}
where $f_\alpha$ is a Dirac spinor built with four test functions belonging to $\mathcal{C}_0^\infty\left(\mathbb{R}^4\right)$. Here, $\bar{f} = f^\dagger \gamma^0$, as usual.  It  is important to remark that this $f_\alpha$, although spinorial, is not anticommuting in nature, being therefore the smeared fields $\psi(f)$ and $\psi^\dagger(f)$ scalar under Lorentz transformations and anticommuting~\cite{Haag:1992hx}. 

The smeared Dirac fields can be rewritten as the sum of the smeared creation and annihilation operators, namely 
\begin{align}\label{key}
	\psi(f) &= b_f + d^\dagger_f, \quad \nonumber \\
	\psi^\dagger(f) &= b^\dagger_f + d_f,
\end{align}
where the smeared operators $b_f, b^\dagger_f, d_f, d^\dagger_f$ can be immediately obtained by plugging Eq.~\eqref{DiracExpansion} into Eq.~\eqref{SmearingDirac}:
\begin{align}\label{key}
	b_f &= \! \int \!\! \frac{d^3k}{(2\pi)^3} \frac{m}{k_0} \sum_{s=1,2} \bar{f}(k) u^{(s)} b_s(k), \nonumber \\
	b_f^\dagger &= \! \int \!\! \frac{d^3k}{(2\pi)^3} \frac{m}{k_0} \sum_{s=1,2} \bar{u}^{(s)}(k) f(k) b_s^\dagger(k), \nonumber \\
	d_f &= \! \int \!\! \frac{d^3k}{(2\pi)^3} \frac{m}{k_0} \sum_{s=1,2} \bar{v}^{(s)}(k) f(-k) d_s(k), \nonumber \\
	d_f^\dagger &= \! \int \!\! \frac{d^3k}{(2\pi)^3} \frac{m}{k_0} \sum_{s=1,2} \bar{f}(-k) v^{(s)}(k) d_s^\dagger(k)\;, 
\end{align}
where $f(k)$ stands for the Fourier transform of $f(x)$: 
\begin{equation} 
f(k) = \int d^4x \; e^{ikx} f(x) \;. \label{FT}
\end{equation} 
The above smeared operators are anticommuting quantities satisfying the following relations:
\begin{align}\label{key}
	\lbrace b_f, b_{f'}^{\dagger} \rbrace &= \int \! \frac{d^3k}{(2\pi)^3} \bar{f}(k) \left(\frac{\slashed{k} + m}{2 \omega_k}\right) f'(k), \nonumber \\
	\lbrace d_g, d_{g'}^{\dagger} \rbrace &= \int \! \frac{d^3k}{(2\pi)^3} \bar{g}'(-k) \left(\frac{\slashed{k} - m}{2 \omega_k}\right) g(-k),
\end{align}
with all the other anticommutation relations vanishing. Moreover, one can always redefine $(b_f,d_g)$ 
\begin{eqnarray} 
b_f & \rightarrow  \left( {\int \! \frac{d^3k}{(2\pi)^3} \bar{f}(k) \left(\frac{\slashed{k} + m}{2 \omega_k}\right) f(k)} \right)^{-1/2} b_f \;, \nonumber \\
d_g & \rightarrow  \left( {\int \! \frac{d^3k}{(2\pi)^3} \bar{g}(-k) \left(\frac{\slashed{k} + m}{2 \omega_k}\right) g(-k)}\right)^{-1/2} d_g \;, \label{redef}
\end{eqnarray} 
so that 
\begin{equation} 
\lbrace b_f, b_{f}^{\dagger} \rbrace = \lbrace d_g, d_{g}^{\dagger} \rbrace = 1 \;, \label{canonic1}
\end{equation} 
which will be taken to hold from now on. 
When acting on the Fock vacuum $\vert 0\rangle$, the smeared operators $(b_f^{\dagger}, d_g^\dagger)$ give rise to states with well-defined finite norms~\cite{Haag:1992hx}. 
\\

\section{Squeezing the vacuum state}\label{SQU}	

In this section we outline the construction of the   Quantum Field Theory version of the GHZ state, building it with smeared fermionic creation and annihilation operators. More precisely, consider the generic smeared fermionic operators $a_f, b_g, c_h, d_l$, where $f,g,h,l$ are smooth test functions in the Schwartz space. Thus, we introduce  the GHZ-like state 
	\begin{align}\label{key}
		\vert \text{GHZ} \rangle^4_\xi = e^{\xi \, a_f^\dagger \, b_g^\dagger \, c_h^\dagger \, d_l^\dagger} \, \vert 0, 0, 0, 0 \rangle,
	\end{align}
where we are considering $\xi \in \left(0,1\right)$. Thanks to the anticommuting nature of the fermionic creation operators, a little algebra shows that 
\begin{align}\label{keyy}
		\vert \text{GHZ} \rangle^4_\xi = \frac{1}{\sqrt{1+\xi^2}} \left(\vert 0, 0, 0, 0 \rangle + \xi \vert 1_f, 1_g, 1_h, 1_l \rangle \right)
\end{align}
Notice that when taking $\xi=1$ we find an expression that is totally analogous to the one considered in the  case of Quantum Mechanics, Eq.~\eqref{key}. It is interesting to note that this GHZ-like state can also be obtained by the application of a {\it squeezing operator} built out with fermionic creation and annihilation operators, defined as
\begin{align}\label{S4}
	S_4(\eta) = e^{\eta \, \left(a_f b_g c_h d_l - a^\dagger_f b^\dagger_g c^\dagger_h d^\dagger_l \right)}
\end{align}  
It is immediate to show that such an operator is unitary, $S_4^\dagger S_4 = S_4 S_4^\dagger = 1$. Notice that, upon acting on the vacuum, this unitary operator yields the GHZ-like state discussed above. In fact, we find
\begin{align}\label{GHZ4}
		\vert \text{GHZ} \rangle^4_\eta &= S_4(\eta) \vert 0 \rangle \nonumber \\
		&= \cos\eta \, \vert 0,0,0,0 \rangle - \sin\eta \, \vert 1_f, 1_g, 1_h, 1_l \rangle.
\end{align}
We remark that these two parametrizations, Eq.\eqref{keyy}  and Eq.~\eqref{GHZ4},  are related to each other by a change of variables of the form $\xi = -\tan\eta$.

Concerning the localization properties of the smeared operators $a_f^\dagger, b_g^\dagger, c_h^\dagger, d_l^\dagger$, as already mentioned in the introduction, we are referring to the Rindler wedges geometry~\cite{Crispino:2007eb,Harlow:2014yka}. Namely, the test functions $f, g, h, l$ are taken as belonging to $\mathcal{C}_0^\infty\left(\mathbb{R}^4\right)$ with their respective supports located in the Rindler wedges space-like separated from each other, according to the principle of relativistic causality. We underline also the fact that the operators $a_f, b_g, c_h, d_l$ anticommute, that is, $	\lbrace a_f, b_g \rbrace = \lbrace a_f, c_h \rbrace = \lbrace a_f, d_l \rbrace = \lbrace b_g, c_k \rbrace = \lbrace b_g, d_l \rbrace  = \lbrace c_h, d_l \rbrace = 0$. As one can easily check, these equations can be obtained by making use of a pair of independent Dirac spinor fields.

Similarly, one can introduce the 3-qubit version of the squeezing operator and the associated GHZ-like state as
\begin{align}\label{GHZ3}
	\vert \text{GHZ} \rangle^3_\eta &= 	S_3(\eta) \vert 0,0,0 \rangle =  e^{i \eta \, \left(a_f b_g c_h  - a_f^\dagger b_g^\dagger c_h^\dagger\right)} \vert 0,0,0 \rangle \nonumber \\
	&= \cos\eta \, \vert 0,0,0 \rangle - i\sin\eta \, \vert 1_f, 1_g, 1_h \rangle,
\end{align}	
where the main difference in comparison with $S_4$ is a factor $i$ in the definition of the $S_3$ operator. 

In order to investigate Mermin's inequalities associated with $M_n$, we introduce a set of $2n$ Hermitian bounded operators that can be defined in the following way: the operator $A_i$ acts only on the {\it i-th} entry, keeping all the others intact, giving a phase $i \alpha_i$ if acts on $\vert0\rangle$, and a phase $-i \alpha_i$ if acts on $\vert1\rangle$. Schematically, considering the non-trivial action of $A_i$ only in the {\it i-th} entry, we have for any $\lbrace x_n \rbrace$:
\begin{align}\label{key}
	A_i \vert x_1,..., x_{i-1}, 0, x_{i+1}, ..., x_n \rangle &= e^{+i \alpha_i} \vert x_1,..., x_{i-1}, 1, x_{i+1}, ..., x_n \rangle, \nonumber \\
	A_i \vert x_1,..., x_{i-1}, 1, x_{i+1}, ..., x_n \rangle   &= e^{-i \alpha_i} \vert x_1,..., x_{i-1}, 0, x_{i+1}, ..., x_n \rangle, 
\end{align}	
with similar expressions for the primed operators. By construction, these Bell-type operators satisfy, as required,  the following properties:
\begin{align}\label{BellOpProp}
	A_i^2 = 1, \quad A_i^\dagger = A_i, \quad \left[A_i, A_j\right] = 0,
\end{align}
with similar equations holding for the primed operators. 

An interesting  feature is that  properties~\eqref{BellOpProp} will also hold for operators obtained from these ones by means of unitary transformations. That is, if we consider transformed operators $\hat{A}_i$ defined as 
\begin{align}\label{key}
	\hat{A}_i = U^\dagger A_i \, U, \quad {\rm with}\quad U^\dagger U = U U^\dagger = 1,
\end{align}
the properties stated in Eq.\eqref{BellOpProp} will keep holding. In particular, we are interested in the unitary transformation defined by the squeezing operator $S(\eta)$.

\section{Mermin's inequalities violation in the Fock vacuum of Quantum Field Theory}\label{M3M4}

The basic equation which will allow us to find maximal violation of Mermin's inequalities in the vacuum state of a relativistic   Quantum Field Theory can be written in a generic form as
\begin{align}\label{Fundamental}
	\langle 0 \vert \hat{M}_n \vert 0 \rangle = \langle 0 \vert S^\dagger M_n \, S \vert 0 \rangle = \langle \text{GHZ} \vert M_n \vert \text{GHZ} \rangle
\end{align}
Equation~\eqref{Fundamental} follows from the invariance properties under unitary transformations of Mermin's operators algebra, Eqs.~\eqref{BellOpProp},~\eqref{key}. We are now ready to investigate the maximum violation of  Mermin's inequalities for $M_3$, $M_4$. 

Let us first consider $M_4$ as defined in Eq.~\eqref{M4}, with operators $\hat{A}, \hat{A}', \hat{B}, \hat{B}', \hat{C}, \hat{C}', \hat{D}, \hat{D'}$, such that $\hat{X} = S_4^\dagger X \, S_4$ for any operator, where $S_4$ is defined as in Eq.~\eqref{S4}. Thus, using the basic equation~\eqref{Fundamental}, we find
\begin{align}\label{key}
	\langle 0 \vert \hat{M}_4 \vert 0 \rangle = \langle 0 \vert S_4^\dagger M_4 \, S_4 \vert 0 \rangle = \langle \text{GHZ} \vert M_4 \vert \text{GHZ} \rangle,
\end{align}
where $\vert \text{GHZ} \rangle$ is given by Eq.~\eqref{GHZ4}. Acting with the $A, B, C, D$ operators on the GHZ-like state will give us a result similar to the one obtained in the   QM case, Eq.~\eqref{M4GHZQM}. In fact, it turns out that upon computing the vacuum expectation value of $\hat{M}_4$ one can obtain

\begin{align}\label{key}
	\langle 0 \vert 2\hat{M}_4 \vert 0 \rangle &= - 2 \, \cos\eta \sin \eta \, \Big[  \nonumber \\ 
	& -\cos (\alpha +\beta +\gamma +\delta ) + \cos (\alpha +\beta +\gamma +\delta') \nonumber \\
	&+ \cos (\alpha +\beta +\gamma'+\delta ) + \cos (\alpha +\beta +\gamma'+\delta')\nonumber \\
	&+\cos (\alpha +\beta'+\gamma +\delta ) + \cos (\alpha +\beta'+\gamma +\delta') \nonumber \\
	&+\cos (\alpha +\beta'+\gamma'+\delta ) - \cos (\alpha +\beta'+\gamma'+\delta') \nonumber \\
	&+\cos (\alpha'+\beta +\gamma +\delta ) + \cos (\alpha'+\beta +\gamma +\delta') \nonumber \\
	&+\cos (\alpha'+\beta +\gamma'+\delta ) - \cos (\alpha'+\beta +\gamma'+\delta') \nonumber \\
	&+ \cos (\alpha'+\beta'+\gamma +\delta ) - \cos (\alpha'+\beta'+\gamma +\delta') \nonumber \\
	&-\cos (\alpha'+\beta'+\gamma'+\delta )-\cos (\alpha'+\beta'+\gamma'+\delta')\Big].
\end{align} 
In order to maximize this expression, we can use the following parameters: $\alpha = 0, \alpha' = \frac{\pi}{2}, \beta = 0, \beta' = \frac{\pi}{2}, \gamma = \frac{\pi}{4}, \gamma' = \frac{3 \pi}{4}, \delta = 0, \delta' = \frac{\pi}{2}, \eta = \frac{\pi}{4}$. We could also have used the parameters adopted to maximize Eq.~\eqref{M4GHZQM}. Therefore, we find in this case the maximal violation:
\begin{align}\label{key}
		\langle 0 \vert 2\hat{M}_4 \vert 0 \rangle = 8 \sqrt{2}.
\end{align}
Performing the same computation for the case $M_3$ defined in Eq.~\eqref{M3}, adopting analogous definitions for the operators $A, A', B, B', C, C'$, see Eq.~\eqref{BellDefABC}, and using the GHZ-like state obtained from the Fock vacuum through the action of the squeezing operator $S_3$~\eqref{GHZ3}, one  finds
\begin{align}\label{key}
	\langle 0 \vert \hat{M}_3 \vert 0 \rangle &= \langle \text{GHZ} \vert M_3 \vert \text{GHZ} \rangle \nonumber \\
	&= -2 \sin \eta \cos \eta \, \Big[ \sin(\alpha' + \beta + \gamma) + \sin(\alpha + \beta' + \gamma) \nonumber \\
	&+ \sin(\alpha + \beta + \gamma') - \sin(\alpha' + \beta' + \gamma') \Big].
\end{align}
To maximize the modulus of this expression, one could take for example: $\alpha = \beta = \gamma = 0, \, \alpha' = \beta' = \gamma' = \frac{\pi}{2}, \, \eta = \frac{\pi}{4}$, giving us the maximal violation $\vert \langle 0 \vert \hat{M}_3 \vert 0 \rangle \vert= 4$. 
A similar computation could be performed with $M_5$, and would give us a totally analogous result: maximum violation.

\section{Conclusions}\label{conc}

In the present work we have presented a Quantum Field Theory  setup for  Mermin's inequalities, which generalize the Bell-CHSH inequality in the case of a $n$-partite system. Our findings are in full agreement with those already obtained by~\cite{Summers:1987fn,Summ,Summers:1987ze,Summers:1988ux,Summers:1995mv} on the Bell-CHSH inequality, being summarized by the following statement: {\it free Dirac quantum field theory leads to the maximum violation of  Mermin's inequalities}. 

This result arises as a consequence of the interplay between various properties which we enlist below: 

\begin{itemize} 

\item The Fock vacuum state $\vert 0\rangle$ of a relativistic   Quantum Field Theory in Minkowski spacetime turns out to be a highly entangled state. This fundamental feature of   Quantum Field Theory  is captured in a very clean way by analyzing the Minkowski spacetime in terms of the Rindler wedges. In doing so, it turns out that the Minkowski vacuum can be expressed as an entangled state in terms of the left and right Rindler modes 
\cite{Crispino:2007eb,Harlow:2014yka,Donnelly:2015hxa,Blommaert:2018rsf,Blommaert:2018oue}. Another way to understand the entanglement properties of the   Quantum Field Theory  vacuum is through the renowned Reeh-Schlieder  theorem, which states that $\vert 0\rangle$ is a cyclic and separating state for the algebra of the local field operators localized in a given open region of the Minkowski spacetime, see~\cite{Witten:2018zva} for an extensive review on this topic. 

\item Mermin's operators $A_i$ entering the $n$-th polynomial $M_n$ fulfill the following algebraic relations: 
\begin{equation} 
A^\dagger_i = A_i\;, \qquad A^2_i = 1 \;,\qquad  \left[ A_i, A_j \right] = 0 \;, \label{Amermin} 
\end{equation} 
which give us the freedom of performing unitary transformations. It is apparent  that the operators 
\begin{equation} 
{\hat A}_i = U\; A_i \; U \;, \qquad U^\dagger\;U = U \; U^\dagger=1 \;, \label{hatAM}
\end{equation}
obey the same algebraic relations, Eq.~\eqref{Amermin}.

\item Use of the smeared Dirac field $\psi(f)$, Eq.~\eqref{SmearingDirac}, in order to introduce the analogue of the GHZ states of Quantum Mechanics. As we have seen, the Dirac spinors seem to be the natural quantum fields to be employed in the  construction of GHZ-like states in   Quantum Field Theory, Eqs.~\eqref{GHZ4},~\eqref{GHZ3}. For instance, Dirac spinors can be employed to construct the squeezing operator 
\begin{align}\label{S4Conc}
	S_4(\eta) = e^{\eta \, \left(a_f b_g c_h d_l - a^\dagger_f b^\dagger_g c^\dagger_h d^\dagger_l \right)}, 
\end{align} 
giving rise to the GHZ state 
\begin{eqnarray}\label{GHZ4conc}
		\vert \text{GHZ} \rangle^4_\eta &= &S_4(\eta) \vert 0 \rangle \nonumber \\
		&= &\cos\eta \, \vert 0,0,0,0 \rangle - \sin\eta \, \vert 1_f, 1_g, 1_h, 1_l \rangle.
\end{eqnarray}

This procedure can be generalized to the case of high order Mermin's polynomials by introducing a suitable multiplet of quantum spinor fields. 

\item Since the GHZ states can be obtained through squeezing unitary operators $S_n$ acting on the vacuum: 
\begin{equation} 
\vert GHZ \rangle_n = S_n \vert 0 \rangle \;, \qquad S_n S_n^\dagger = S_n^\dagger S_n = 1 \;, \label{sqconc}
\end{equation} 
it follows that, see Eq.~\eqref{Fundamental}, 
\begin{equation}\label{Fundconc}
	\langle 0 \vert \hat{M}_n \vert 0 \rangle = \langle 0 \vert S_n^\dagger M_n \, S_n \vert 0 \rangle = \langle \text{GHZ} \vert M_n \vert \text{GHZ} \rangle,
\end{equation}
where $\hat{M}_n$ and $M_n$ are related by the unitary transformation $S_n$, Eq.~\eqref{sqconc}. This equation has a rather simple and interesting meaning: it relates the expectation value of $\hat{M}_n$ in the vacuum state $\vert0\rangle$ to the expectation value of the unitary equivalent operator $M_n$ evaluated in the GHZ state. As we have already seen, Eq.\eqref{Fundconc} leads to the maximum violation of  Mermin's inequalities. 
\\

\item 
We end the conclusion with a final remark on our understanding of the vacuum state in Quantum Field Theory. As already stated, we rely on the Rindler wedges geometry~\cite{Crispino:2007eb}, enabling us to express the vacuum state $\vert 0 \rangle$ as an entangled state in terms of the left and right modes, namely~\cite{Crispino:2007eb}:
\begin{align}\label{key}
	\vert 0 \rangle = \prod_{i} \left[ C_i \sum_{n_i=0}^{\infty} e^{-\frac{\pi n_i \omega_i}{a}} \, \vert n_i, R \rangle \otimes \vert n_i, L \rangle   \right],
\end{align}
where $C_i = \sqrt{1 - \exp\left(-2 \pi \omega_i/a\right)}$.
One notices that the above expression depends on the so-called Rindler parameter $a$, deeply related with the Unruh temperature~\cite{Crispino:2007eb}. Since the vacuum state is cyclic, this dependence affects the whole Hilbert space of Quantum Field Theory. Due to the intrinsic presence of the parameter $a$, we are allowed to use the dimensionless variable $\frac{a}{m}$. As such, the parameter $\eta$ entering the unitary operator $S_n$ expression in Eq.~\eqref{S4Conc} is thought to be a function of $\frac{a}{m}$, meaning that we are not introducing any kind of new extra hidden parameter. We are just taking advantage of the intrinsic parameter $a$, which expresses the entangled nature of the vacuum in Quantum Field Theory. Said otherwise, in a Quantum Field Theory, it is possible to perform a Bell-CHSH test in the vacuum state itself by making use of unitary transformations, as expressed by Eq.~\eqref{Fundamental}. 
Of course, this is only possible because the vacuum state of Quantum Field Theory is entangled. 

\end{itemize} 


\begin{acknowledgments}
	The authors would like to thank the Brazilian agencies CNPq and FAPERJ, for financial support. S.P.~Sorella  and  I.~Roditi are both CNPq researchers, respectively under the contracts 301030/2019-7 and  311876/2021-8.
\end{acknowledgments}


\end{document}